\documentclass[10pt,conference]{IEEEtran}
\IEEEoverridecommandlockouts
\usepackage{cite}
\usepackage{amsmath,amssymb,amsfonts}
\usepackage{algorithmic}
\usepackage{graphicx}
\usepackage{textcomp}
\usepackage{xcolor}
\usepackage{caption}
\usepackage{tabu}
\usepackage{svg}
\usepackage{url}
\usepackage{tabularx}
\usepackage{blindtext}
\usepackage{subcaption}
\usepackage{longtable}
\usepackage{array} % For m{width} column type
\usepackage{hyperref}
\usepackage{booktabs}

\usepackage{color, colortbl}
\usepackage{adjustbox}
\definecolor{Gray}{gray}{0.9}

\usepackage{tikz}

\title{Transforming Fashion with AI: A Comparative Study of Large Language Models in Apparel Design}
\author{
    \IEEEauthorblockN{Nusrat Jahan Lamia\IEEEauthorrefmark{1}, Sadia Afrin Mim \IEEEauthorrefmark{2}} 
    \IEEEauthorblockA{\IEEEauthorrefmark{1}Bangladesh University of Textiles}\IEEEauthorblockA{\IEEEauthorrefmark{2} George Mason University}

    \IEEEauthorrefmark{1}njlamia17@gmail.com
    \IEEEauthorrefmark{2}safrinmi@gmu.edu
    }

\usepackage{bibentry}
\begin{document}

\maketitle

\begin{abstract}
Fashion has evolved from handcrafted designs to automated production over the years, where AI has added another dimension to it. Nowadays, practically every industry uses artificial models to automate their operations. To explore their role, we examined three prominent LLMs (OpenAI, GeminiAI, Deepseek) in multiple stages of textile manufacturing (e.g., sustainable choice, cost effectiveness, production planning, etc.). We assessed the models' ability to replicate garment design using certain parameters (fabric construction, shade, weave, silhouette, etc.). We compared the models in terms of different body types and functional purposes (e.g., fashionwear, sportswear) so that designers could evaluate effectiveness before developing actual patterns, make necessary modifications, and conduct fashion forecasting beforehand. To facilitate deeper analysis, we created a custom dataset specifically for fabric image generation and classification. Our analysis revealed that, in terms of fabric construction, the OpenAI DALL-E model integrated with ChatGPT outperformed other models, achieving a lower LPIPS (Learned Perceptual Image Patch Similarity) score of approximately $0.2$. In fabric classification from images, we found OpenAI offered the best results by breaking down certain factors (e.g., breathability, moisture-wicking, and tactile comfort), achieving approximately $80\%$ accuracy for base construction and $55\%$ for detailed construction. However, our results indicate that Deepseek faced significant challenges in generating and recognizing fabric images. Overall, all the models struggled to recognize complex fabric constructions and intricate designs from images, and relying too much on AI might hinder human creativity. We also observed that all three models performed effectively in providing recommendations and insights for fabric design in textual form.

% 50\% bias mitigate 32% active
\end{abstract}
\section{Introduction}

\footnote{This paper is accepted on \textbf{3rd International Conference on Textile Science and Engineering}}
For a very long time, fashion has been an embodiment of craftsmanship, innovation and civilization. In the past, it required a lot of effort to design and manufacture garments which largely depended on skilled artisans who meticulously crafted each item by hand. With the advancement of technology, automation has entered the process and production is being streamlined by digital pattern making, advanced industrial sewing machines, and computer aided design (CAD) integration with computer aided manufacturing (CAM). However, the most recent change in the garment industry which is the adoption of Artificial intelligence (AI), is influencing the production process to a large extent from design selection to trend forecasting. Adopting AI in the garment industry could be greatly beneficial to streamline manufacturing, improve production efficiency and forecast fashion with greater accuracy. Large Language Models (LLMs) and Generative AI are automating critical stages, from material selection and style generation to production and trend analysis. This integration improves efficiency and sustainability, enabling personalized fashion solutions. Tools like DALL-E, Gemini, and Deepseek are streamlining processes, assisting with sample making, cost analysis, and virtual try-on. Designers are now collaborating with AI to predict trends and optimize fabric choices. While AI enhances productivity and innovation, questions persist about its capacity to fully replicate or surpass human creativity. The balance between AI's analytical power and human artistic intuition remains a key consideration in the future of fashion.
Several research studies have explored AI’s role in fashion. Monteresso et al examined how AI models contribute to the fashion design process highlighting how AI can optimize creation processes, improve operational efficiency and shorten development times without compromising the quality and creativity[1]. Preliminary analysis revealed that ChatGPT and The New Black AI are the most suited tools. Qianqian Chen et al introduced Fashion-GPT, a model suitable for fashion retrieval systems and compared its performance with ChatGPT, Gemini and Deepseek[2]. Their findings showed that Fashion-GPT outperformed general models like ChatGPT in fashion-specific tasks particularly multi-view, multi-modal matching in fashion queries. Wenda Shi provided an overview which discussed incorporation of GenAI into fashion-related tasks and applications, multi-modal fashion understanding and fashion synthesis of image[3] with 53 publicly available datasets suitable for training and benchmarking fashion-centric models. Han Liu et al focused on "Sequential LLM Framework for Fashion Recommendation" which revealed that while most e-commerce recommendation systems are not fashion-specific, a new mix-up-based retrieval technique significantly enhances product recommendations[4]. Manish Kumar Meeshra et al. explored the role of AI chatbots in the fashion industry which demonstrated how AI-driven chatbots, utilized by Natural Language Processing (NLP) and deep learning models such as RNN, Seq2Seq, LSTM, BERT and GPT enhance customer experiences by providing personalized recommendations, order tracking and efficient customer support[5]. Yoon Kyun Lee et al. incorporated research using ChatGPT-3.5 AND ChatGPT-4,  compared AI-generated predictions with real-world and generated good enough results based on user input[6]. Geetha Manoharan et al. examined AI’s impact on fashion supply chain optimization, demonstrating how AI-driven models enhance stock management by employing AI models trained on past inventory and sales data to reduce waste and enhance marketing strategies[7]. Anastasiia Movchaniuk et al. explored the use of Leonardo AI for clothing collection design, generating and refining women’s jacket designs based on classic and romantic styles by using different neural networks to conceptualize clothing collections[8]. Esmeralda Marku et al. analyzed AI’s role in enhancing designer creativity, automating the design process and overcoming the algorithmic limitations to enhance and automate their creative process for handling complex fashion problems[9]. Victoria Rodriguez Schon et al. worked on AI’s impact on trend research within fashion, evaluating its capacity for ethical and nuanced analysis in fashion trend research, emphasizing the need for responsible AI utilization that respects cultural and social contexts for a thoughtful approach beyond mere replicating trends[10]. Risa Aihara et al. compared ChatGPT generated fashion recommendations to those made by university students for the similar fashion items, highlighting AI’s limitations or strengths in image recognition and random item selection[11]. Finally, Eleonora Pantano et al. examined Generative AI’s role in luxury fashion design, which revealed that consumers perceived that the GAI-designed luxury products reflect and reinforce the essence and symbolic values of the brands while varying in perceived creativity and manufacturing quality[12].

\section{Methodology}
\label{method}

%  \begin{figure*}[ht]
%     \centering
%     \includegraphics[width=\textwidth]{CameraReady/LaTeX/aequitas-language.png}
%     \caption{The `Languages' information provided by GitHub for the \textsc{Aequitas} intervention repository.}
%     \label{fig:languages}
% \end{figure*}

% \begin{figure}
%     \centering
%     \includegraphics[width=2in]{CameraReady/LaTeX/lift-about.png}
%     \caption{Information regarding the mode of use for \textsc{LiFT} provided in the `About'.}
%     \label{fig:lift-about}
% \end{figure}

% \begin{figure}
%     \centering
%     \includegraphics[width=2in]{CameraReady/LaTeX/fairlearn-license.png}
%     \caption{License information provided in the \textsc{fairlearn} repository.}
%     \label{fig:fairlearn-license}
% \end{figure}

%\samim{Typically I recommend starting the methodology section by re-iterating the goal of the research (e.g., ``The goal of our research is to...'' and then outlining the RQs that guided the work (e.g., ``To accomplish this goal, we collected and analyzed data to answer the following research questions:)}
We aim to map and analyze current large language models (LLMs) to determine how they can be effectively used in the fashion industry.
We gathered and examined intervention name-specific data to answer the following research questions:

\begin{description}
    % \item[\textbf{RQ1}] \textit{What ecosystem do open source fairness interventions offer to practitioners and developers?}
    % \samim{what do we mean by ecosystem? Could we say something like compatibility instead?}
    \item[\textbf{RQ1}] \textit{How well do these LLMs perform tasks that are typically associated with fashion design?
?}
    \item[\textbf{RQ2}] \textit{What are the challenges for these models for design  construction?}

\end{description}

% To answer these research questions, we used a previously compiled list of interventions~\cite{mim2023taxonomy} to systematically identify and expand our dataset to include 52 more open source interventions.
 % \item How wide are these interventions in terms of application? 
% how applicable are these in different contexts. models and domains they apply to
% jegula pai an otao finding a add kora jay
%\item What kind of ML algorithmic fairness do these interventions work on? 

% \item What are the availability of these interventions and how are the intervention communities trying to support these interventions? 
% - language support
% - openness
% - platform/mode of use
% - licensing

% - operation availability
%\item How customizable are these interventions for data points?

% \end{itemize}

%\subsection{RQ5 Analysis}
%To analyze RQ5 we analyzed
%\begin{itemize}
 %   \item Data point customization ability which is indicated in the data point edit access node in both the preproces and postprocess phases. We find out what kind of customization these interventions offer by reading through the README files, or web demos, and tutorials. For example for PAIR-code/what-if-intervention we found in their tutorial that this intervention allows pre-modification of data before running a model. Similarly for dssg/aequitas we found in their web demo that these interventions allow post-modification of parameters to adjust results.
%\end{itemize}

\subsection{Study Design}
This study explores the effectiveness of various Large Language Models (LLMs) in performing essential tasks within the fashion industry, including design replication, material identification, and production planning. To support this evaluation, we compiled a curated dataset consisting of diverse fabric images, featuring textiles such as cotton, wool, linen, satin, mesh, lace, flannel, georgette, pique, polyester and its blends, tweed, and corduroy, among others. The dataset also included detailed descriptions capturing specific garment elements like buttons, sleeves, necklines, and embroidery patterns.
For the credibility of our analysis, we incorporated fabric images that had been previously annotated by textile industry practitioners and publicly available on the web. Additionally, to assess the fabric classification capability of the LLMs, we utilized images from leading online retailers taking advantage of the models’ internet access to retrieve and process the content. Beyond that, we introduced a set of unique, industry-specific fabric images that were not publicly available, ensuring their authenticity through verification by experienced textile professionals. In total, our dataset comprised 100 clothing images collected from online retailing sources (Shein , Amazon , H\&M) \& from Esquire Knit Composite Ltd. factory. Images were in a diverse format (JPEG, PNG) with an average resolution of 1024 × 1024 pixels. We selected these images based on three criteria: (1) the presence of detailed fabric texture and construction, (2) the inclusion of visible garment elements such as sleeves, necklines, or embellishments, and (3) availability of textual product descriptions that could support verification by textile industry practitioners. The selection aimed to ensure diversity in clothing categories (e.g., casualwear, formalwear, activewear) and fabric types (e.g., cotton, satin, velvet).
Furthermore, we developed 50 descriptive prompts through an explorative listing  process by our authors. Our authors found a few product titles from online exploration from Google Shopping and the details. Our goal was to simulate realistic design requests by incorporating a wide range of attributes, including fabric types (e.g., velvet, cotton blends, mesh), color palettes (including Pantone references), body prints (e.g., floral, abstract, character-based), button specifications (e.g., snap, enamel, shank), and neckline styles (e.g., Mandarin, Queen Anne, V-neck). The prompts were intentionally  mixed within designs to vary in complexity.

\subsection{LLM Model Selection  \&  Evaluation}

For evaluation with LLM models we considered 3 most popular LLM models used worldwide which are: Gemini AI, OpenAI,  DeepSeek. We also included another LLM model, FashionGPT, in our evaluation. Unlike the other LLM models, FashionGPT is not freely available. However, we assessed its performance alongside the others to a certain limit to gain a comprehensive understanding of its capabilities in fashion-related tasks. The models were tested on the following criteria:
Fabric identification accuracy: by comparing them to the authentic labels, which were the expert-verified classifications of the fabric images, we assessed if the LLM models could accurately identify fabric type and construction or not.
Level of detail in fabric, embellishment or finishing classification: Evaluating AI classifications' specificity (e.g., differentiating cotton varieties, finish type, wash type, type of embellishments used etc).
Sample design creation: Evaluating if the models could generate sample design according to given requirements.
Material recommendations – Providing insights on production based on certain parameters such as climate suitability, sustainability, comfort and durability.
To ensure reliable results, each LLM was tested with the same set of images and similar queries multiple times. The models were given standardized questions to evaluate their material and other (fabric, embellishment, finishing) classification and alternative suggestion abilities. For FashionGPT, we couldn’t find the image generation option. It could perform some specific tasks like outfit suggestion, fashion combination, color \& fabric coordination, styling tips, fashion advice etc which could eventually be done  by the three LLM models we are discussing. 

\section{Findings}
In this section, we present and analyze the LLM models’ performance in key fashion related tasks: Image classification, Image generation, Style recommendation. The results are analyzed to evaluate the efficiency, accuracy and effectiveness of these models.
\subsection{Image Classification}
\begin{figure*}[ht]
  \includegraphics[width=15cm]{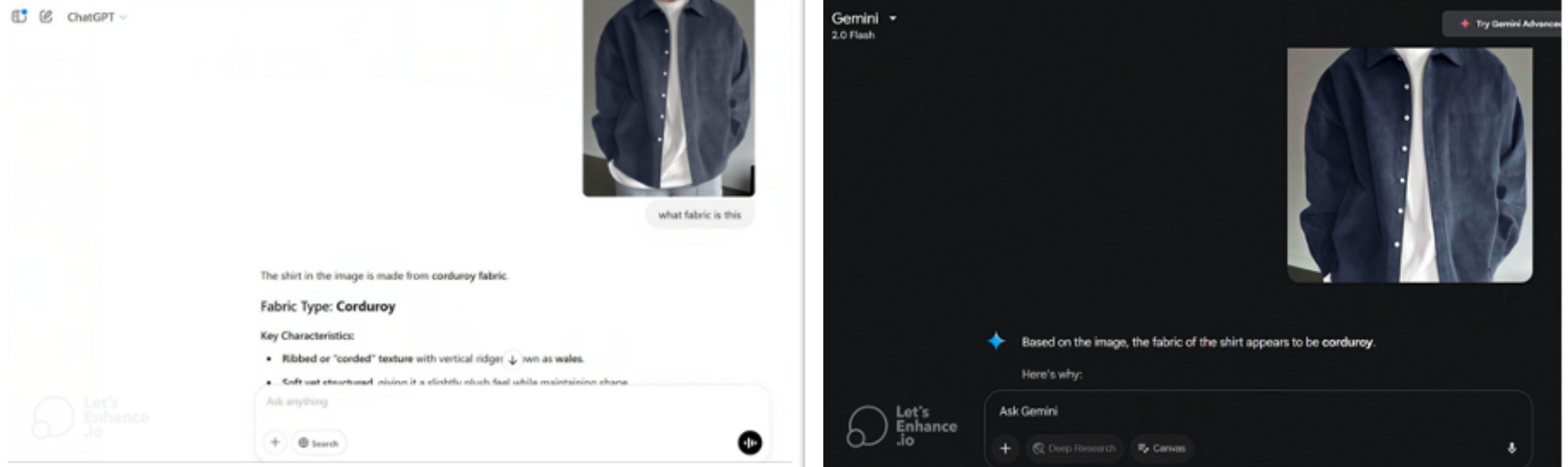}\centering
  \caption{ Input-1 classified by OpenAI’s DALL-E      VS      Input-1 classified by GeminiAI’s Imagen 3} 
  \label{classify1}
\end{figure*}

\begin{figure*}[ht]
  \includegraphics[width=15cm]{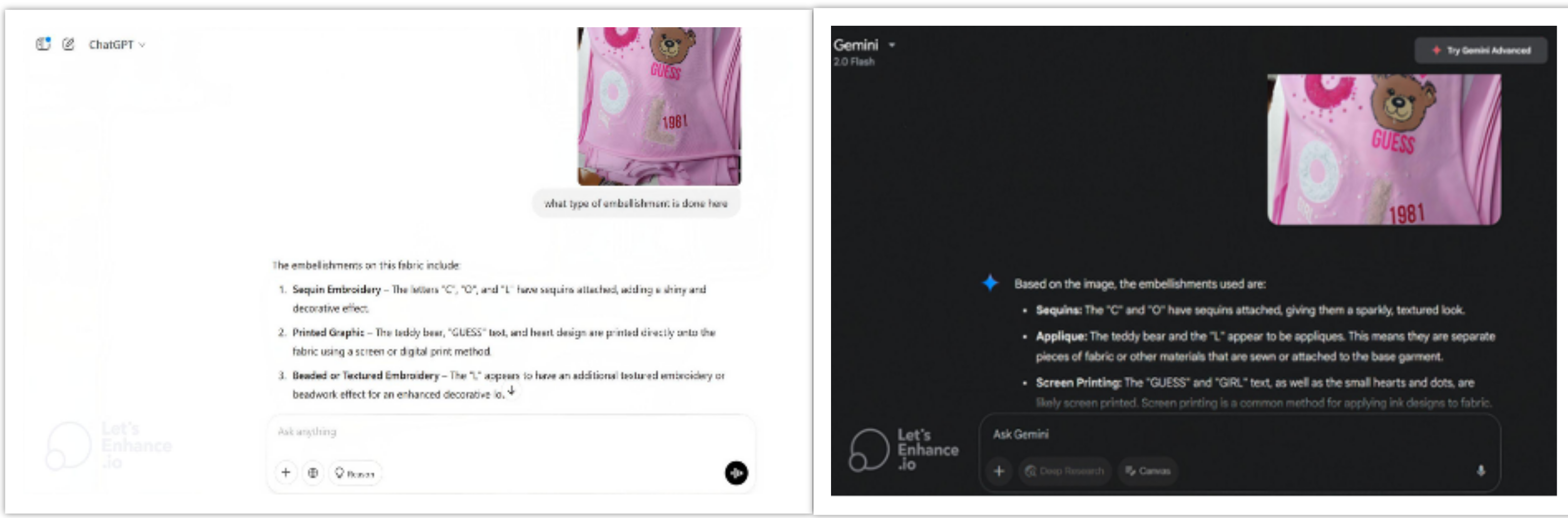}\centering
  \caption{ Input-2 classified by OpenAI’s DALL-E    VS      Input-2 classified by GeminiAI’s Imagen 3} 
  \label{classify2}
\end{figure*}

\begin{figure*}[ht]
  \includegraphics[width=15cm]{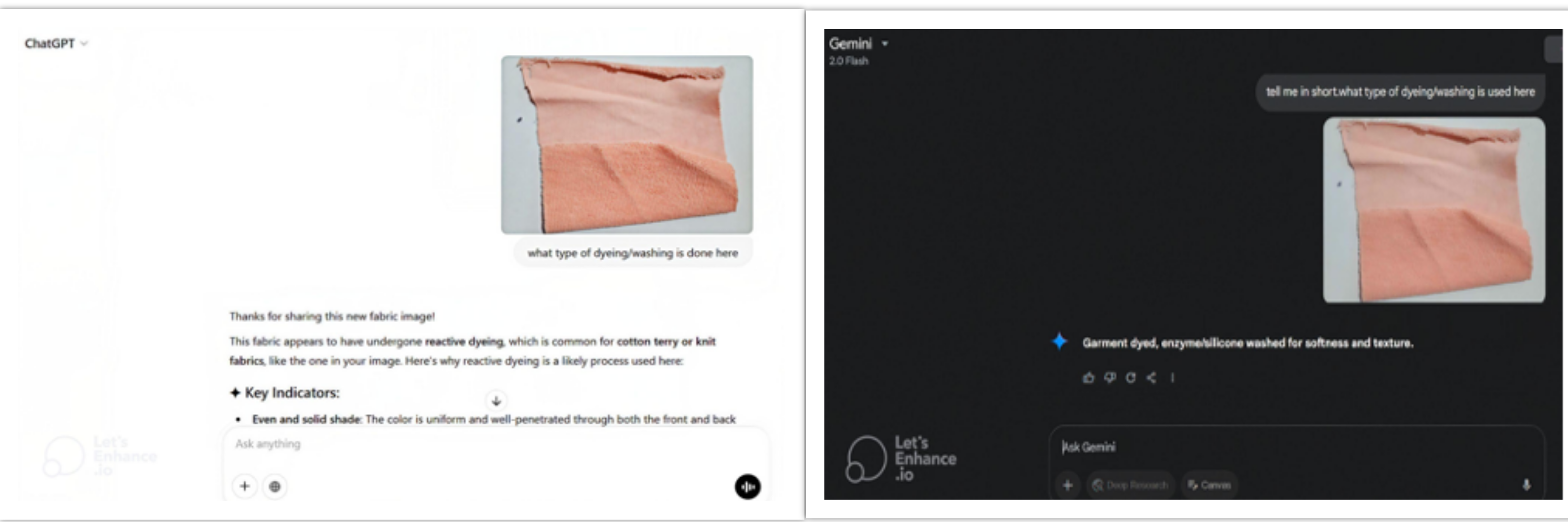}\centering
  \caption{Input-3 classified by OpenAI’s DALL-E  VS     Input-3 classified by GeminiAI’s Imagen 3} 
  \label{classify3}
\end{figure*}

We aim to identify the ability of LLM models to classify different fabric types and construction, embellishment and finishes based on image inputs. We are showing here some samples of our experiment. 
\begin{itemize}
    \item Figure~\ref{classify1} shows the prompt for input-1 and the textile industry practitioner confirmed the generated result to be accurate for both OpenAI model \& Gemini Model.
    \item Figure~\ref{classify2} shows the prompt for input-2 and the textile industry practitioner confirmed the generated result to be accurate for OpenAI. However, for Gemini the identification of sequin and screen printing was correct but the applique  part was incorrect as rubber printing was used here.
    \item Figure~\ref{classify3} shows the prompt for input-3. While OpenAI classified reactive dyeing, our textile industry practitioner concluded that pigment dyeing was used here. Although the fabric classification was correct, which is terry fabric. For Gemini, Our textile industry practitioner confirmed the identification of Garment dyeing procedure was correct but the wash type identified was incorrect.
\end{itemize}
\subsubsection{Key Insights from Image Classification:} 
\begin{figure}[ht]
  \includegraphics[width=8cm]{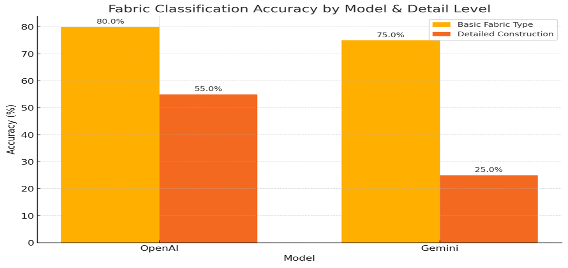}\centering
  \caption{ Comparison of LLM models for fabric image classification} 
  \label{result1}
\end{figure}

OpenAI outperformed Gemini in identifying fabric construction, embellishment and finishing. Although OpenAI couldn’t identify them all properly compared with actual results, still the results were quite satisfactory as shown in Figure~\ref{result1}.  The figure indicates the percentage of  LLM models classifying the fabric image. Therefore the basic labels include fabric type and color, detailed construction include button type, neckline, sleeves. Their performance is quite similar with OpenAI showing slightly advanced performance both in terms of basic \& complex fabric recognition which 80\% \& 55\% accordingly. While DeepSeek is still a new LLM model, it needs further development. Till now it can not directly analyze images so we couldn’t conduct the image classification test for DeepSeek. Also  FashionGPT, not having sufficient features, couldn’t analyze the given images.
\subsection{Image Generation}
\begin{figure*}[ht]
  \includegraphics[width=15cm]{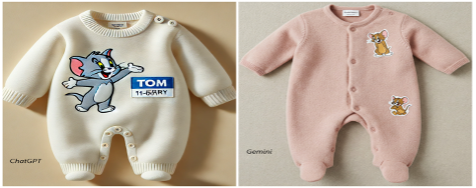}\centering
  \caption{ Image generated by OpenAI’s DALL-E VS  Gemin’s Imagen 3 for design description-1} 
  \label{gen1}
\end{figure*}

\begin{figure*}[ht]
  \includegraphics[width=15cm]{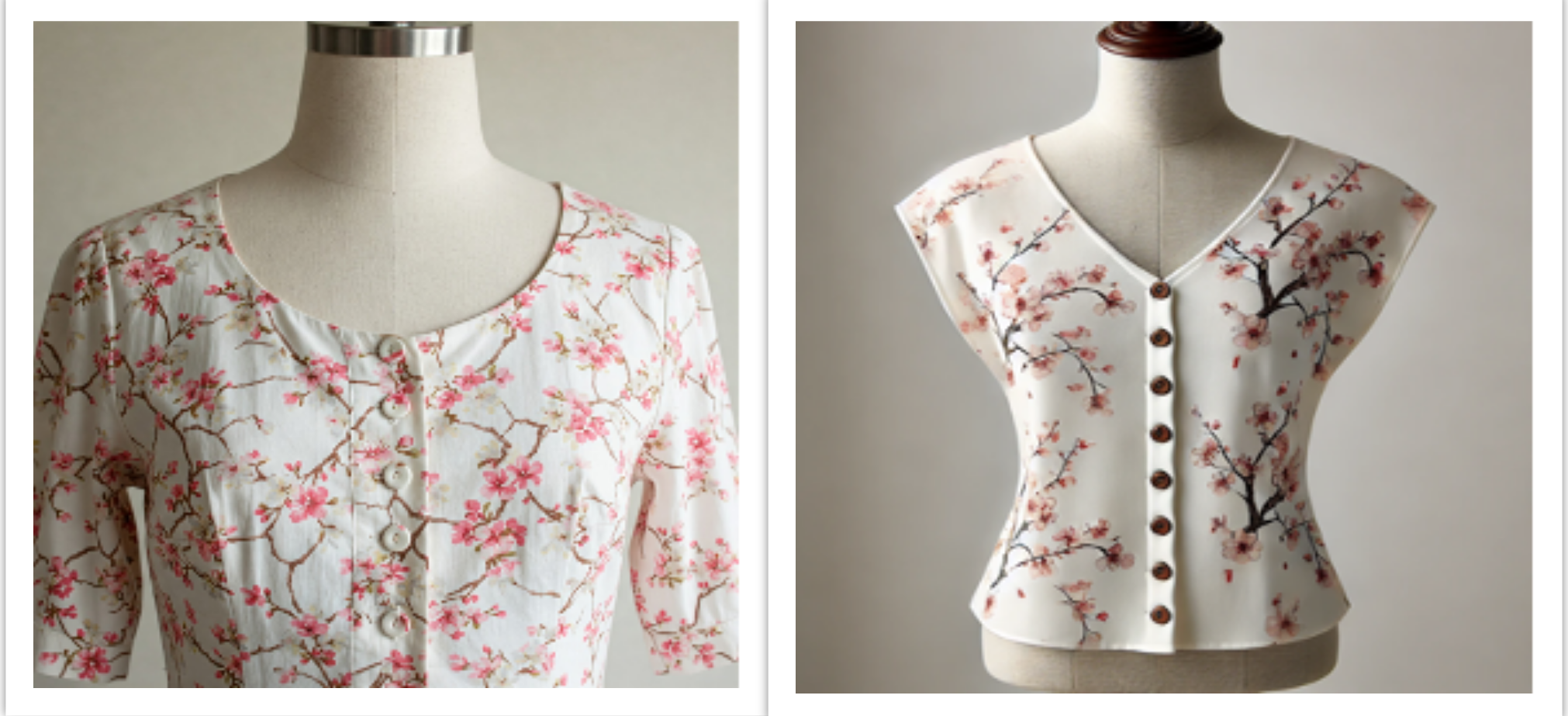}\centering
  \caption{Image generated by OpenAI’s DALL-E VS  Gemini’s Imagen 3 for design description-2} 
  \label{gen2}
\end{figure*}

\begin{figure*}[ht]
  \includegraphics[width=15cm]{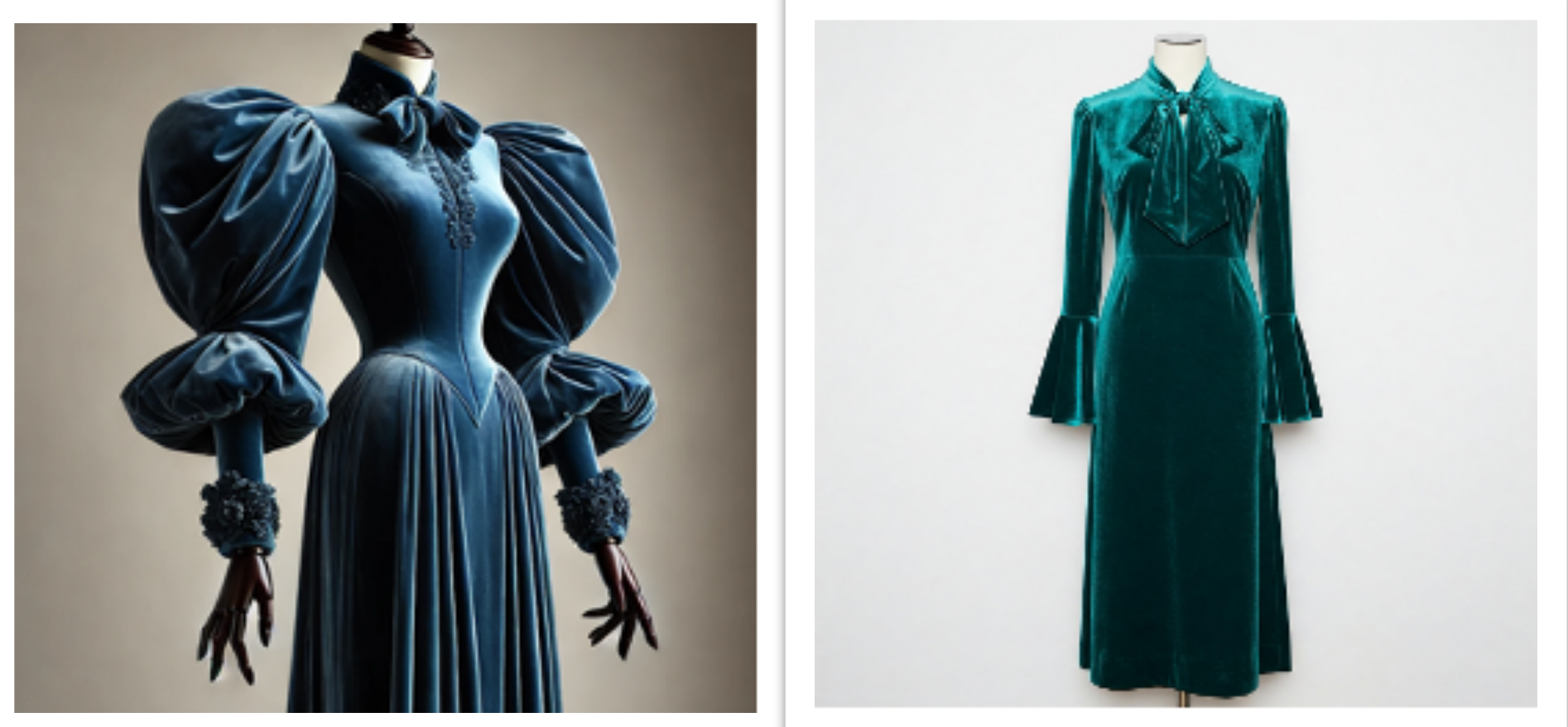}\centering
  \caption{Image generated by OpenAI’s DALL-E VS  Gemini’s Imagen 3 for design description-3} 
  \label{gen3}
\end{figure*}

\begin{figure*}[ht]
  \includegraphics[width=15cm]{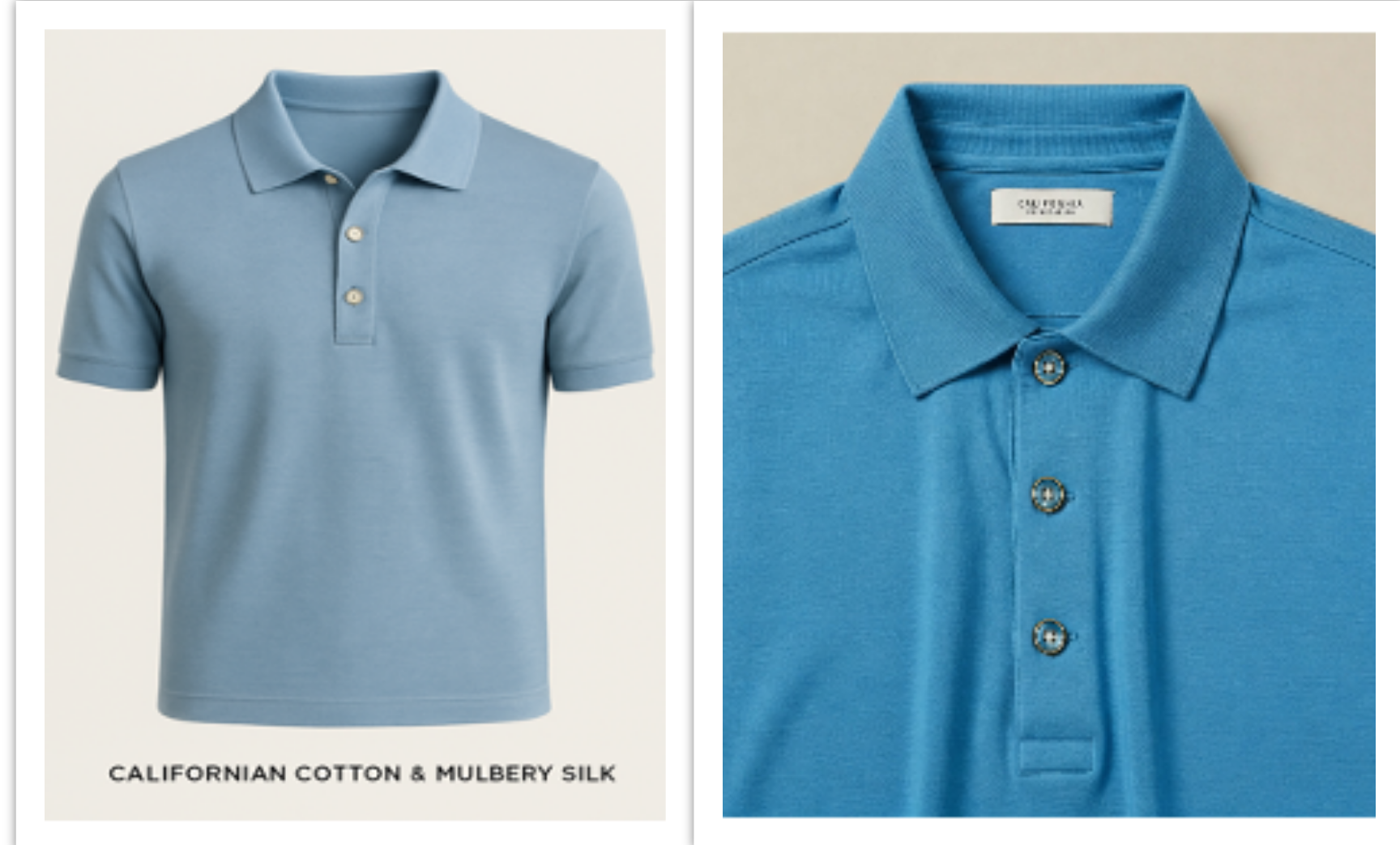}\centering
  \caption{Image generated by OpenAI’s DALL-E VS  Gemini’s Imagen 3 for design description-4} 
  \label{gen4}
\end{figure*}

From our dataset of design descriptions for generating images we showed here some samples for the LLM models generating images for different clothes.
\begin{itemize}
    \item For design description-1 the prompt is “Create an image for a sample baby clothing suitable for winter with cashmere wool, pantone color 11-0601 TCX, tom and jerry sticker print on chest, snap button (25 mm dia) on sleeve, lining on the inside and let me know the additional details you used.”  The generated result is shown in Figure~\ref{gen1}.\\
    \textbf{Observation:} After observing the generated images we found that: OpenAI’s DALL-E generated the sample design according to given instruction but it didn’t use snap buttons, rather seems to have used polyester buttons. Gemini AI’s Imagen 3 generated the sample design without following the given pantone color code. Also it seems to have used decorative or fabric buttons. 

    \item For design description-2 the prompt is “generate image for a clothing item with Shank buttons, Queen Anne neck, White color, Cherry blossom print” shown in Figure~\ref{gen2}.\\
    \textbf{Observation:} Upon reviewing the generated images, we observed that OpenAI’s DALL-E model was able to produce sample designs based on the provided instructions; however, it failed to incorporate specific elements such as Shank buttons and the Queen Anne neckline. Similarly, Gemini AI’s Imagen 3 model generated the sample designs without accurately rendering the button details. Nevertheless, both models demonstrated strong performance in replicating color schemes and body prints.

   \item For design description-3 the prompt is “Cyan, Velvet, Gigot Sleeve Dress, Mandarin Collar, Bow Tie long dress” shown in Figure~\ref{gen3}.\\
   \textbf{Observation:} OpenAI’s DALL-E model generated the sample design following the provided instructions; however, it did not accurately replicate the specified cyan color. In contrast, Gemini AI’s Imagen 3 model produced a design that closely adhered to the given instructions.

    \item For design description-4 the prompt is “Generate an image for Californian Cotton \& Mulberry Silk sky blue polo shirt for men with Enamel Buttons” shown in Figure~\ref{gen4}.\\
\textbf{Observation:} OpenAI’s DALL·E model generated the sample design in alignment with the provided instructions but omitted the use of Enamel buttons. Similarly, Gemini AI’s Imagen 3 followed the instructions but failed to accurately depict the specified button type. Additionally, both models produced varying shades of sky blue, with Gemini’s output leaning towards a noticeably darker shade.

\end{itemize}

\subsubsection{Key Insights from Image Generation:} 
\begin{figure*}[ht]
  \includegraphics[width=15cm]{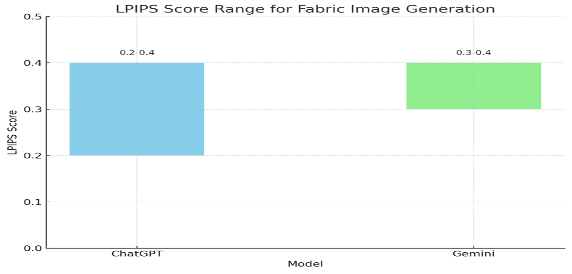}\centering
  \caption{Comparison of LLM models for fabric image generation
} 
  \label{result2}
\end{figure*}

Our findings indicate that both OpenAI and Gemini AI generated fabric images that closely resemble human creativity. For generating Figure~\ref{result2}, we collected ground truth images corresponding to our prompts from Google Shopping, they reflected realistic clothing. We then generated corresponding images using the selected LLM models. To evaluate the perceptual similarity between the generated outputs and their respective references, we calculated the LPIPS scores by installing and utilizing the official LPIPS Python package. This allowed us to quantitatively assess the visual fidelity of each model's image generation performance.  The closer the  LPIPS score to 0 the  higher visual similarity between the images. As illustrated in Figure 9, the LPIPS scores for the images produced by both models suggest a strong alignment with human design aesthetics. Notably, OpenAI demonstrated slightly superior performance in this aspect, although its results exhibited greater variability compared to Gemini AI. Text based AI models such as FashionGPT and DeepSeek offer conceptual guidance rather than visual outputs. Improvement in AI generated texture, draping and detailing is needed for practical application in fashion design.

\subsection{Style Recommendation
}
The objective of this kind of prompt generation is  to evaluate how well AI models recommend outfits and styles based on given preferences. We provided a similar scenario to the three LLM models and analyzed the accuracy and efficiency of their answers for a similar prompt, “Suggest me an outfit for upcoming Eid-Ul-Fitr which will be in summer. Suggest something that will be comfortable and pleasing to look at.” Open AI ChatGPT and GeminiAI both suggested outfits with suitable fabric and trendy colors for the given scenario along with additional styling details shown in Figure 10 \& 11. Although in this case DeepSeek showed additional insights on suggesting sustainable style and fabric for both male and female shown in Figure 12. FashionGPT, being a specialized model for fashion purposes, outperformed the three models by recommending styles for men, women and unisex wear.

\begin{figure*}[ht]
  \includegraphics[width=15cm]{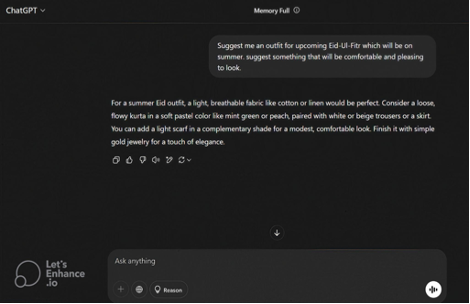}\centering
  \caption{Recommendation by OpenAI’s DALL-E } 
  \label{reco1}
\end{figure*}
\begin{figure*}[ht]
  \includegraphics[width=15cm]{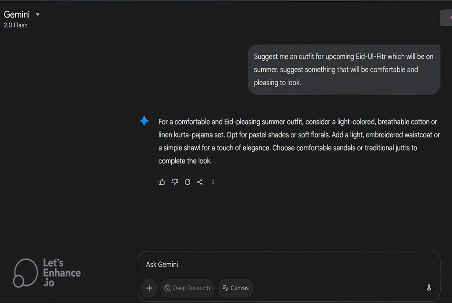}\centering
  \caption{Recommendation by  Gemini’s Imagen 3
} 
  \label{reco2}
\end{figure*}
\begin{figure*}[ht]
  \includegraphics[width=15cm]{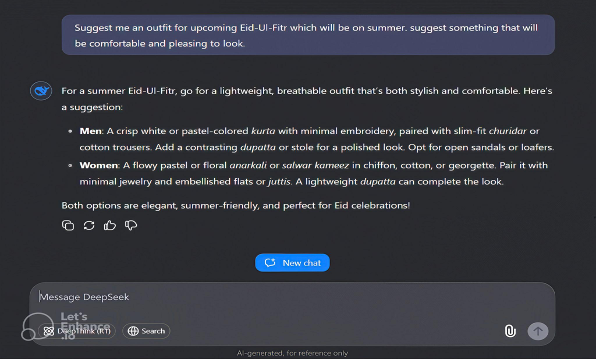}\centering
  \caption{Recommendation by Deepseek} 
  \label{reco3}
\end{figure*}

\section{Conclusion}
Analyzing the results and discussions, it is clear that LLM models have a lot of potential to improve the fashion industry. The models showed supremacy in fabric classification achieving up to 80\% accuracy in OpenAI LLM model although there were some problems in identifying unusual materials \& subtle details (e.g.,button, neckline, pantone color, wash type etc). Regarding image generation, OpenAI and Gemini created impressive sample designs ranging LPIPS scores 0.2 to 0.4 based on given inputs although there were some lackings in human-like detailing and innovation. For style recommendation, all three of the  models did a good job of suggesting outfits based on current trends, sustainability and functionality. While each model has its strengths and weaknesses, they displayed that AI has the potential to contribute in fashion sectors that no one imagined before.

\end{document}